\documentclass[a4paper,12pt]{article}

\usepackage{amsmath,amssymb}
\usepackage{array}
\usepackage[nospace]{cite}
\usepackage{mcite}
\usepackage[dvips]{graphicx}
\usepackage{a4p}
\usepackage{l3_title}

\newcommand{\Zob}{\ensuremath{\mathrm{Z}}}
\newcommand{\epem}{\ensuremath{\mathrm{e}^+\mathrm{e}^-}}
\newcommand{\elpr}{\ensuremath{\mathrm{ep}}}
\newcommand{\qqbar}{\ensuremath{\mathrm{q}\bar\mathrm{q}}}
\newcommand{\pion}[1]{\ensuremath{\pi^{#1}}}
\newcommand{\pipmpipm}{\pion{\pm}\pion{\pm}}
\newcommand{\pipmpimp}{\pion{\pm}\pion{\mp}}
\newcommand{\piopio}{\pion{0}\pion{0}}
\newcommand{\ra}{\ensuremath{\rightarrow}}
\newcommand{\GeV}{\ifmmode {\mathrm{\ Ge\kern -0.1em V}}\else
                            \textrm{~Ge\kern -0.1em V}\fi}%
\newcommand{\MeV}{\ifmmode {\mathrm{\ Me\kern -0.1em V}}\else
                            \textrm{~Me\kern -0.1em V}\fi}%
\newcommand{\fm}{\ifmmode  {\mathrm{\ fm}}\else
                            \textrm{~fm}\fi}
\newcommand{\mm}{\ifmmode  {\mathrm{\ mm}}\else
                            \textrm{~mm}\fi}
\newcommand{\jetset}{\textsc{Jetset}}
\newcommand{\jetsetbe}{\textsc{Jetset-BE}}
\newcommand{\jetsetnobe}{\textsc{Jetset-noBE}}
\newcommand{\luboei}{\texttt{LUBOEI}}
\newcommand{\etal}{{\itshape et~al.}}
\newcommand{\ie}{{\itshape i.e.}}
\newcommand{\eg}{{\itshape e.g.}}
\newcommand{\Ebump}{\ensuremath{E_{\text{cluster}}}}
\newcommand{\Epio}{\ensuremath{E_{\gamma\gamma}}}
\newcommand{\Etrack}{\ensuremath{E_{\text{track}}}}
\newcommand{\minv}{\ensuremath{m_{\gamma\gamma}}}
\newcommand{\fbg}[1]{\ensuremath{f_{\text{bg}}#1}}
\newcommand{\fsig}[1]{\ensuremath{f_{\pi}#1}}
\newcommand{\ftwod}[1]{\ensuremath{f_{\text{2d}}#1}}
\newcommand{\fbgbg}[1]{\ensuremath{f_{\text{bgbg}}#1}}
\newcommand{\Apipi}[1]{\ensuremath{A_{\pi\pi}#1}}
\newcommand{\Apibg}[1]{\ensuremath{A_{\pi\text{bg}}#1}}
\newcommand{\Abgbg}[1]{\ensuremath{A_{\text{bgbg}}#1}}
\newcommand{\cq}{\ensuremath{R_2(Q)}}

\journalname{Phys. Lett. B}
\date{31 August 2001}
\preprint{2001-063}

\begin{document}

\begin{titlepage}

\title{Bose-Einstein Correlations of Neutral and Charged
       Pions in Hadronic Z Decays} 
\author{The L3 Collaboration}

\begin{abstract}
  Bose-Einstein correlations of both neutral and like-sign charged
  pion pairs are measured in a sample of 2 million hadronic \Zob{}
  decays collected with the L3 detector at LEP. The analysis is
  performed in the four-momentum difference range $300\MeV < Q <
  2\GeV$. The radius of the neutral pion source is found to be smaller
  than that of charged pions. This result is in qualitative agreement
  with the string fragmentation model.
\end{abstract}

\submitted

\end{titlepage}

\section*{Introduction}

Bose-Einstein correlations (BEC), between identical bosons, have been
extensively studied in hadronic final states produced in \epem, \elpr,
hadron-hadron and heavy-ion interactions
\cite{Goldhaber:1959,Boal:1990,Baym:1998}. The bosons studied are
mainly charged pions 
\cite{L3_174,Aleph_91.4}.
Only rarely have neutral pions been studied
\cite{Eskreys:1969,Grishin:1988}, and never before in \epem{}
interactions.  In this letter, we report a study of BEC of \pion{0}
pairs in hadronic decays of the \Zob{} boson at LEP, and compare them to
BEC of pairs of identically charged pions.

BEC manifest themselves as an enhanced production of pairs of identical
bosons which are close to one another in phase space. This can be studied
in terms of the two-particle correlation function $R_2$ \cite{L3_174}:
\begin{equation} 
R_2(p_1,p_2) = \frac{\rho_{\text{BE}}(p_1, p_2)}{\rho_0(p_1,p_2)}, 
\label{eq:C_2}
\end{equation}
where $\rho_{\text{BE}}(p_1, p_2)$ is the two-particle number density for
identical bosons with four-momenta $p_1$ and $p_2$, subject to
Bose-Einstein symmetry.  The reference distribution, $\rho_0(p_1,p_2)$,
is the same density in the absence of Bose-Einstein symmetry.

Assuming a static spherical boson source with a Gaussian density and a plane
wave description of the bosons, $R_2(p_1,p_2)$ is written as
\cite{GGLP:1960,Deutschmann:1982}:
\begin{equation}
\cq =  {\cal N} (1 + \alpha Q) (1 + \lambda e^{ - Q^2 R^2 } ),
\label{eq:C_gausexp}
\end{equation}
where $Q^2 = - (p_1-p_2)^2$ is the square of the four-momentum
difference. The parameter $R$ can be interpreted as the size of the
boson source 
in the centre-of-mass system of the boson pair
and a measurement of the correlation function $R_2$ gives
access to the source size.  The parameter $\lambda$ is introduced to
describe the fraction of effectively interfering pion pairs.  In this
analysis the normalization factor ${\cal N}(1 + \alpha Q)$ is added.
It takes into account possible long-range momentum correlations, as
well as possible differences in pion multiplicity in the data and
reference samples, over the four-momentum difference range studied.

The spherical shape of the boson source assumed here is a simplified
picture. High statistics charged pion data at LEP revealed the source
to be elongated \cite{L3_174,Delphi_234}.  The present
measurement has, however, no sensitivity to the shape of the source of
neutral pions, due to its limited statistics.

Several theoretical predictions exist for differences in BEC for pairs
of bosons in the pion isospin triplet (\pion{+}, \pion{-}, \pion{0}).
From the string model \cite{Andersson:1983} a smaller spatial emission
region, \ie\ a wider momentum correlation distribution is expected for
\piopio{}, than for \pipmpipm{}. This follows from the break-up of the
string into \qqbar{} pairs, which forbids two equally charged pions to
lie next to each other on the string, whereas two neutral pions can.
The same effect is found when the probabilistic string break-up rule
is interpreted as the square of a quantum mechanical amplitude
\cite{Andersson:1986,Andersson:1998}. From a quantum statistical
approach to Bose-Einstein symmetry \cite{Andreev:1991}, a small
difference between \pipmpipm{} and \piopio{} correlation is expected.
The size and shape of this difference is predicted to be similar to an
expected Bose-Einstein correlation of \pipmpimp{} pairs.
It is theoretically uncertain whether BEC between unlike sign pions are
also to be expected on the ground of isospin invariance 
\cite{Suzuki:1987}.

The main purpose of this letter is to measure the difference 
in size of the emission region of neutral and charged pions. In order to
minimize systematic uncertainties on this difference, the procedures
followed for the charged pions are kept as close as possible to those
for the neutral pions. 

\section*{Data and Monte Carlo Samples}

For this analysis, data collected with the L3 detector
\cite{L3_00} during the 1994 and
1995 LEP runs are used. The analysis is based mainly on measurements
from the high resolution electromagnetic calorimeter and from the
central tracking device.  The data sample corresponds to an integrated
luminosity of about $78\ \mathrm{pb}^{-1}$ at centre-of-mass energies
around $\sqrt{s} = 91.2\GeV$.  From this sample, about 2 million
hadronic events are selected, using energy deposits in the
electromagnetic and hadronic calorimeters \cite{L3_202}.

The \jetset{} generator \cite{Sjostrand:1994} is used to study the
detector response to hadronic events.  Parameters of the generator are
tuned to give a good description of event and jet shapes of hadronic
events measured in L3.  The effects of Bose-Einstein symmetry are
simulated with the \luboei{} routine \cite{Lonnblad:1995}.  
The routine has two parameters, which have been chosen to obtain a 
reasonable description of L3 data. 
This
\textit{ad-hoc} model shifts boson momenta after the hadronization
phase in such a way that the correlation function \cq{} for identical
bosons is proportional to a constant plus a Gaussian as in 
Equation~\ref{eq:C_gausexp}.  
The generated events are passed
through a full detector simulation \cite{L3_SIL3} and are
reconstructed and subjected to the same analysis procedure as the
data.  This Monte Carlo sample (\jetsetbe{}) contains about 7 million
events.  A control sample (\jetsetnobe{}) is also generated with
\jetset{} but it has \luboei{} switched off and the generator
parameters retuned.  Significant differences are found in the tuned
parameters in the two cases. The number of events in the control
sample is approximately 2 million.  Unless stated otherwise, the
\jetsetbe{} sample is used throughout this letter.

\subsection*{Neutral Pion Selection}

Neutral pions in hadronic events are reconstructed from photon pairs.
Photon candidates are identified in the electromagnetic calorimeter as
a cluster of at least two adjacent crystals. The clusters are required
to be located in the central region of the detector,
$|\cos(\theta_{\text{cluster}}) | < 0.73$, and to be in the energy
range $100\MeV < \Ebump < 6 \GeV$.  Above 6\GeV{}, the two photons
from a \pion{0} decay can no longer be distinguished as two separate
clusters.

Discrimination of clusters originating from photons or electrons from
those due to other particles is based on the distribution of the
energy over the crystals of the cluster. A good photon discrimination
is achieved with a neural network based on this energy distribution
\cite{Sanders:2002}.
To reject clusters due to charged particles, a minimum distance
between the cluster and the extrapolation of the closest track of
30\mm{} is required. This corresponds to 1.5 times the size of the
front face of a crystal.

Pairs of photon candidates within an event are used to reconstruct
\pion{0}'s. The distribution of invariant mass of photon pairs shows a
peak around the \pion{0} mass, above a smooth background.  These two
components are extracted by a fit to the mass spectrum. The background
is described by a Chebyshev polynomial of third order, \fbg{}.  The
\pion{0} peak is parameterized by a Gaussian function with exponential
tails, which is continuous, and smooth in the first derivative:
\begin{equation}
\fsig{(\minv)} = 
  \left\{ 
  \begin{array}{l@{\quad\text{if}\quad}l} 
     \exp\left(  \frac{\alpha^2}{2}\right) 
     \exp\left(  \frac{\alpha(\minv-m_\pi)}{\sigma} \right)
     & \minv - m_\pi < -\alpha\sigma  \\
     \exp\left( -\frac{(\minv-m_\pi)^2}{2 \sigma^2} \right) 
     & \beta\sigma \geq \minv-m_\pi \geq -\alpha\sigma  \\
     \exp\left(  \frac{\beta^2}{2} \right) 
     \exp\left( -\frac{\beta(\minv-m_\pi)}{\sigma}  \right)
     & \minv - m_\pi >  \beta\sigma   \\
  \end{array}
  \right. .
\label{eq:fsig}
\end{equation}
Here, \minv{} is the two-photon invariant mass, $m_\pi$ indicates the peak
position, and $\sigma$ is the width of the Gaussian peak. The
parameters $\alpha$ and $\beta$ determine the values of \minv{} where
the Gaussian changes into an exponential.  The
exponential tail at low mass is needed to describe the presence of
converted photons in cluster pairs. If a photon converts, \eg, in the
outer wall of the tracker, it can still be selected as a photon,
although some of the original photon energy is lost.  The high mass
tail accounts for an overestimate of the cluster energy due to another
cluster nearby.

\begin{figure}[htb]
\begin{center}
\includegraphics[width=.707\textwidth]{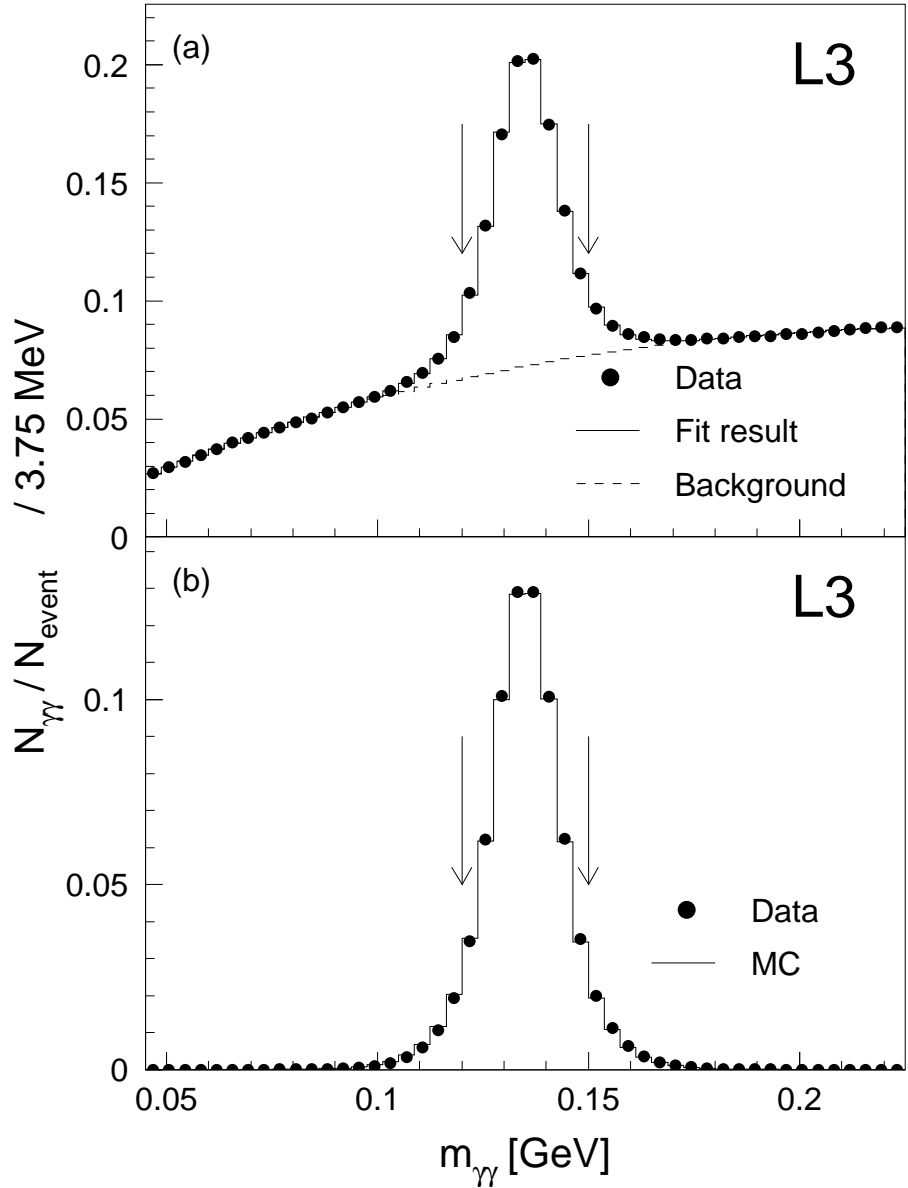}
\caption{Distribution of (a) the two-photon invariant mass \minv{} for
  data together with the fit result and (b) the \pion{0} signal
  as obtained from
  fits to data and to Monte Carlo. The arrows indicate the 
  mass selection window.
  \label{fig:minv1d} }
\end{center}
\end{figure}

An example of the fit result is given in Figure~\ref{fig:minv1d}. The
sum of the two functions \fbg{} and \fsig{} describes the distribution
well. Also, the shape and the size of the \pion{0} peak in the data
and Monte Carlo agree. The mass resolution as determined from the fit
is about $7.4 \MeV$.

The photon pair is then selected as a \pion{0} candidate if it has an
energy $200 \MeV < \Epio < 6 \GeV$.  For a mass window $120 \MeV <
\minv < 150 \MeV$, a total of $1.3$ million \pion{0}'s is selected in
data. The \pion{0} purity of the candidate sample is of the order of
$54 \%$ and the \pion{0} selection efficiency is approximately $17 \%$.
About half of the background is combinatorial, \ie{}, photons from different 
\pion{0}'s. The other half consists of pairs where one or both of the 
photons do not come from a \pion{0} decay.

\subsection*{Charged Pion Selection}

Charged pions are detected as tracks in the central tracker. They are
selected in the same kinematic range as neutral pions:
$|\cos(\theta_{\text{track}}) | < 0.73$ and $200\MeV < \Etrack < 6
\GeV$, where the energy is calculated from the track momentum,
assuming the \pion{\pm} mass. In addition, at least 35 out of 62
possible wire hits are required in the track fit, and the number of
wires between the first and the last must be at least 50.
Furthermore, the track must have at least one hit in the inner part of
the tracker, and the distance of closest approach to the \epem{}
vertex in the plane transverse to the beam is required to be less than
5\mm{}. Finally a high resolution measurement of the polar angle is
demanded. Charged pions are analyzed in the 1995 data only, in which
$4.1$ million tracks are selected.

\section*{Pion Pair Analysis}
\subsection*{Neutral Pions}

\begin{figure}[htb]
\begin{center}
\includegraphics[width=\textwidth]{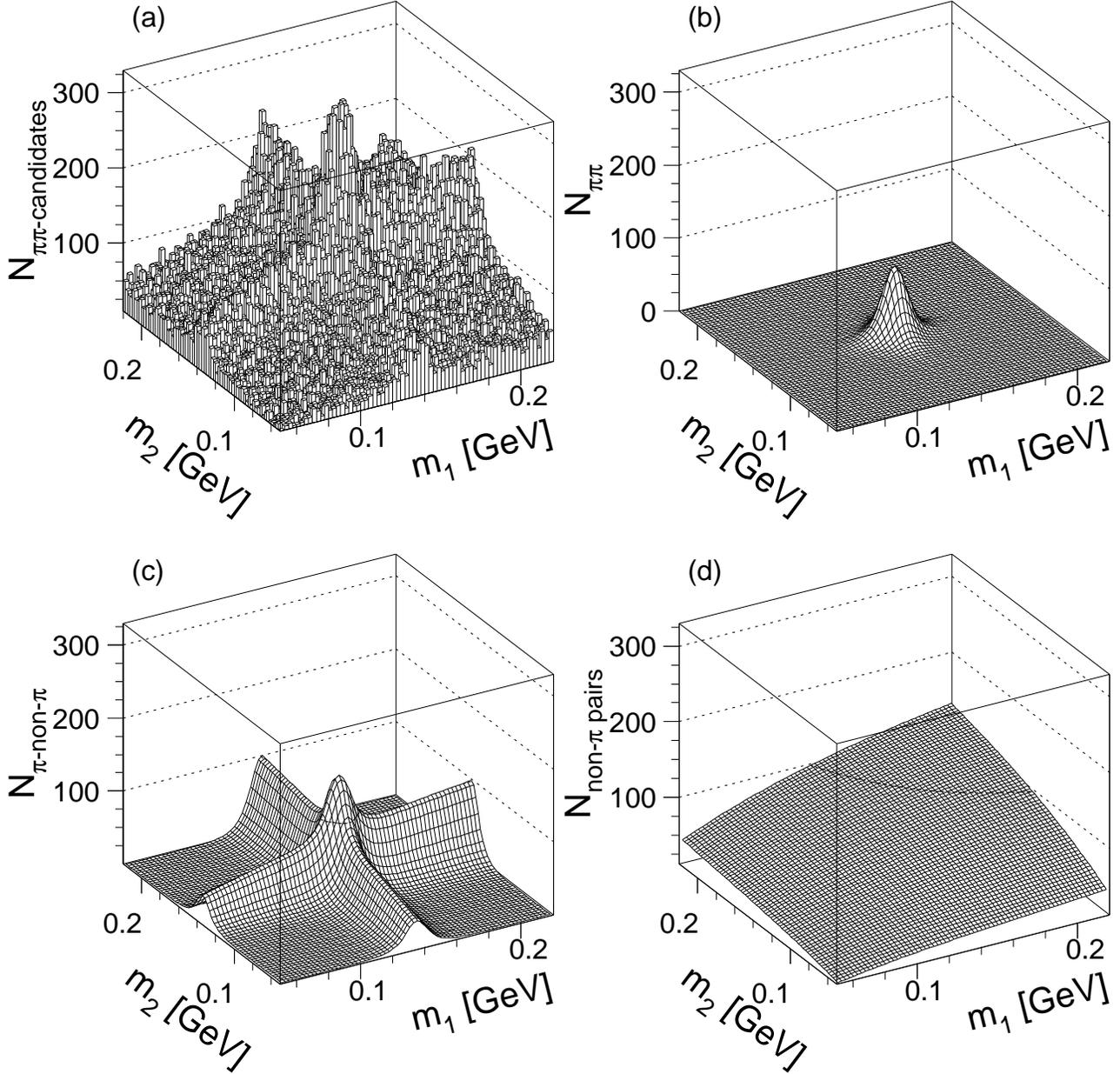}
\caption{(a) Two-dimensional distribution of the mass of \pion{0} pair
  candidates with a four-momentum difference in the range $0.48 < Q <
  0.52 \GeV$, in bins of $2.5 \times 2.5 \MeV^2$. Result of the fit
  for (b) \pion{0} pairs, (c) \pion{0} with
  non-\pion{0} pairs, (d) non-\pion{0} pairs.
  \label{fig:minv2d} }
\end{center}
\end{figure}

After the neutral pion selection, \pion{0} candidates within an event
are paired, requiring that no cluster is common to the two
candidates, and their four-momentum difference $Q$ is calculated.  The
\piopio{} component of the $Q$ distribution is estimated by a fit to
the two-dimensional mass spectrum for every bin in $Q$.  An example of
these mass distributions is shown in Figure~\ref{fig:minv2d}a, for
the bin $0.48 < Q < 0.52 \GeV$. The various contributions are clearly
visible: non-\pion{0} pairs give the smooth background, \pion{0} with
non-\pion{0} pairs give the two ``ridges'' in the \pion{0} peak regions,
\pion{0} pairs give part of the peak in the centre of the plot, the
other part being caused by the sum of the \pion{0} with non-\pion{0}
ridges.

This two-dimensional distribution is derived from the product of two
one-dimensional mass distributions \cite{Sanders:2002}:
\begin{equation}\begin{split}
\ftwod{(m_1,m_2)} & = \Apipi \fsig{(m_1)} \fsig{(m_2)} \\
  & + \Apibg [ \fsig{(m_1)} \fbg{(m_2)} +
                    \fbg{(m_1)} \fsig{(m_2)} ] \\
  & + \Abgbg \fbgbg{(m_1,m_2)} ,
\label{eq:f2d}
\end{split}\end{equation}
where the first term describes the \piopio{} part, the second term
describes the \pion{0} with non-\pion{0} pair ridges and the third
term is the non-\pion{0} pair background. The number of \pion{0}
pairs in the mass window $120\MeV < m_{1,2} < 150\MeV$ is related to the
parameter \Apipi{}.  The functions \fsig{} and \fbg{} have the same
functional form as described before, \fbgbg{} follows from the product
of two Chebyshev polynomials of third order and is required to be
symmetric in the two masses.  All 18 parameters of \ftwod{} are left
free in the fit.

The result of a binned maximum likelihood fit for the mass
distribution in Figure~\ref{fig:minv2d}a is shown in
Figures~\ref{fig:minv2d}b--d. In this representative example, the
$\chi^2$ is 4737 for 4606 degrees of freedom, which corresponds to a $9
\%$ confidence level.

\begin{figure}[htbp]
\begin{center}
\includegraphics[width=.707\textwidth]{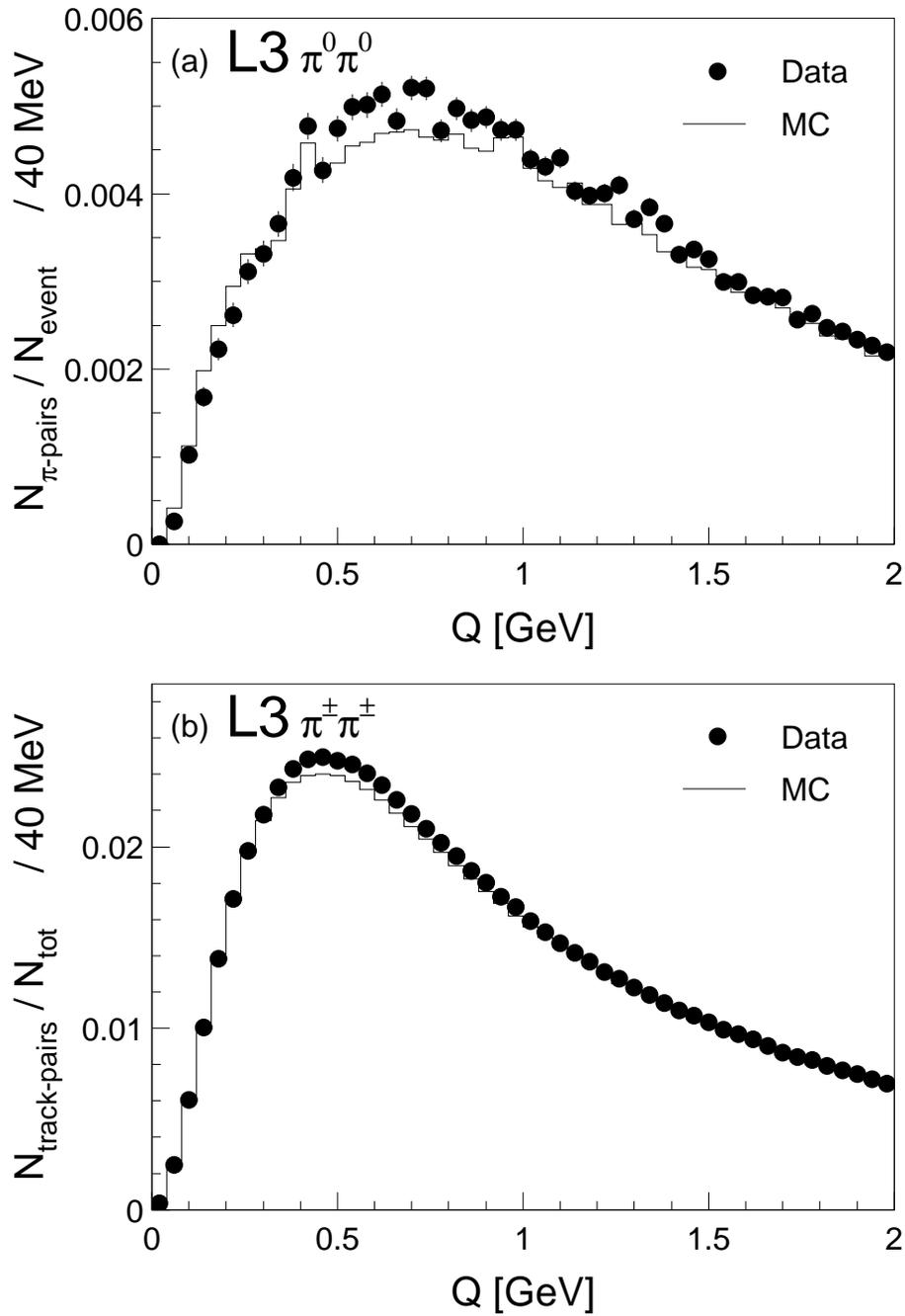}
\caption{Data and Monte Carlo distribution of the four-momentum
  difference of (a) pairs of \pion{0}'s, as obtained from fits of 
  Equation~\ref{eq:f2d}, and
  (b) pairs of \pion{\pm} candidates.
  \label{fig:pionpair_Q} }
\end{center}
\end{figure}

Figure~\ref{fig:pionpair_Q}a presents the $Q$ distribution for
\pion{0} pairs in the mass window $120\MeV < m_{1,2} < 150\MeV$,
obtained from the
values of the parameter \Apipi{} from two-dimensional mass fits of 
Equation~\ref{eq:f2d} to both data and Monte
Carlo. Some deviations between data and Monte Carlo are caused by the
imperfect modeling of BEC. The efficiency to select a \pion{0} pair in
an event ranges from about $1 \%$ at $Q=300\MeV$ to $4 \%$ at
$Q=2\GeV$.

Bins in $Q$ below 300\MeV{} are not used for the rest of the analysis
for two reasons.  First, the efficiency estimate depends strongly on
the BEC modeling in the generator, in the region of small $Q$.
This occurs because the BEC modeling moves identical pions closer 
together, which lowers the detection efficiency.
Secondly, the four-momentum difference of any pair of \pion{0}'s from
$\eta \ra \piopio\pion{0}$ decays is kinematically constrained to have
$Q < 311.7 \MeV$, and in that $Q$-range, more than $20 \%$ of all
\pion{0} pairs originate from this decay.  A rejection of the small
$Q$ region thus avoids systematic uncertainties due to the simulation
of the $\eta$ multiplicity.

\subsection*{Charged Pions}

The distribution of four-momentum difference of equally charged pion
pairs is obtained by calculating $Q$ for pairs of tracks selected
within an event and with the same charge.  This raw \pipmpipm{}
spectrum is corrected bin-by-bin for both pion purity and efficiency
using the Monte Carlo simulation.  The uncorrected distribution is
shown in Figure~\ref{fig:pionpair_Q}b.  Compared to the \piopio{}
case, smaller deviations are observed between the raw spectrum in data
and Monte Carlo. These deviations are due to the imperfect modeling of
BEC in the Monte Carlo.

\section*{Results}
\subsection*{Neutral Pions}

To obtain the final correlation function \cq{}, the 
$Q$ distribution of \pion{0} pairs 
in the data, Figure~\ref{fig:pionpair_Q}a, is
corrected for selection efficiencies. The efficiency is defined as
the number of selected \pion{0} pairs in Monte Carlo events
(\jetsetbe) divided by the number of generated \pion{0} pairs in the
same events, where the generated pions have to be in the same
kinematic range as the selected pions. This definition includes an
acceptance correction for those \pion{0}'s which cannot be selected
kinematically. In this way, the correlations of \pion{0} pairs can
directly be compared to those of charged pion pairs.  The reference
distribution $\rho_0(Q)$ is calculated from a \jetsetnobe\ sample at
generator level, where pions are taken in the same kinematic range as
in the definition of the selection efficiency. 
We choose this reference distribution rather than the distribution for
\pipmpimp{} because of the uncertainty concerning BEC between unlike
sign pions mentioned in the introduction. 
The correlation
distribution \cq{} is then the ratio of the corrected data spectrum to
the reference spectrum.

\begin{figure}[htb]
\begin{center}
\includegraphics[width=.707\textwidth]{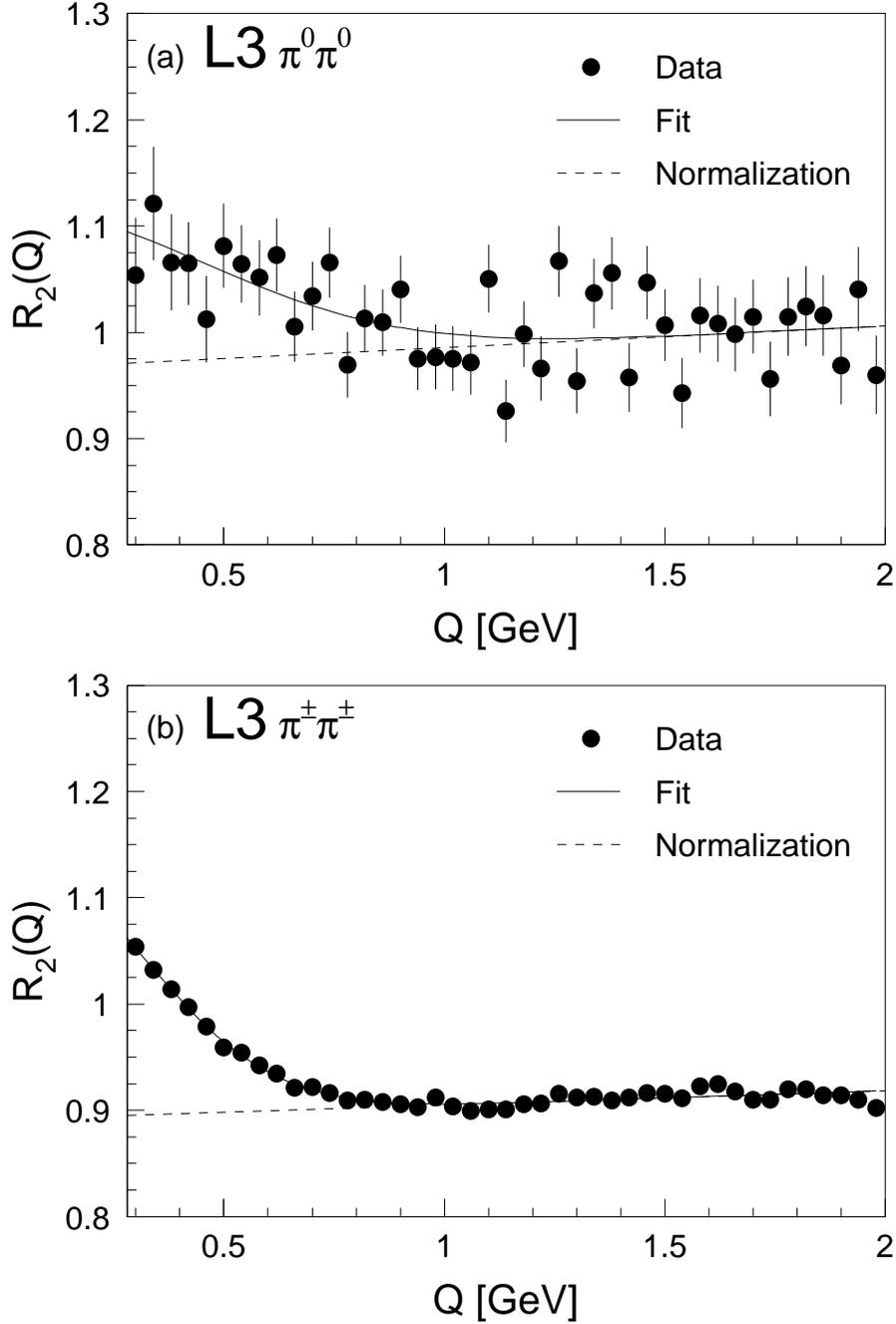}
\caption{Distribution of \cq{} for (a) \piopio{} and (b) \pipmpipm{},
  and results of the fits.  The points indicate the data, the full
  line corresponds to the fit result and the dashed line is the
  normalization factor ${\cal N}(1+\alpha Q)$.
  \label{fig:pion_BE} }
\end{center}
\end{figure}

The distribution of \cq{} is displayed in Figure~\ref{fig:pion_BE}a.
An enhancement at low $Q$ values, expected from Bose-Einstein
symmetry, is clearly visible.  The function \cq{} from
Equation~\ref{eq:C_gausexp} is fitted to this ratio
in the interval $300 \MeV < Q < 2 \GeV$. Extending the fit to lower 
values of $Q$ results in consistent values of the parameters but with
much larger systematic uncertainties. 
The overall
normalization ${\cal N}$ is determined from the integrals of \cq{} and
the fit function; the only free parameters are $\lambda$, $R$ and
$\alpha$.  In this fit, the $\chi^2$ is 46.1 for 40 degrees of
freedom, corresponding to a $23 \%$ confidence level.

The systematic uncertainty on the result due to the \pion{0} selection
is determined by varying the photon selection cuts and by changing the
size of the \pion{0} mass window by $\pm 10 \MeV$. The \piopio{} mass
fit of Equation~\ref{eq:f2d}
is tested by varying the fit range by $\pm 12.5 \MeV$. The
uncertainty due to the modeling of Bose-Einstein correlations in the
Monte Carlo generator is taken into account by using the control
sample \jetsetnobe{} in the efficiency correction procedure. In
addition, the influence on the final result of the agreement
between data and Monte Carlo of distributions relevant to 
the photon and \pion{0} selection,
such as neural network output and energy and polar angle of \pion{0}'s,
is studied. Finally, the binning in $Q$ is varied.  The
systematic uncertainty on the result due to each of the sources, is
assigned as half the maximum deviation.  A summary is given in
Table~\ref{tab:pi0syst}.  The total systematic uncertainty is
calculated as the quadratic sum of these uncertainties.

\subsection*{Charged Pions}

The final correlation function \cq{} for \pipmpipm{} is calculated in
a similar way as that for \piopio{}. The \pipmpipm{} data
distribution, Figure~\ref{fig:pionpair_Q}b, is corrected for purity and
efficiency.  As for \piopio{}, the efficiency is calculated for
generated pions in the same kinematic range as the selected pions.
The reference distribution $\rho_0(Q)$ and the correlation distribution
\cq{} are obtained in the same way as for \piopio{}.

The correlation function for \pipmpipm{} is shown in
Figure~\ref{fig:pion_BE}b.  Due to the higher selection efficiency for
charged pions as compared to neutral pions, the significance of the
low $Q$ value enhancement is much larger. As for \piopio{}, the
function defined in Equation~\ref{eq:C_gausexp} is fitted to the
final distribution. In this fit, the $\chi^2$ is 42.6 for 40 degrees
of freedom, corresponding to a $36 \%$ confidence level.

The systematic uncertainty on the result due to the track selection is
determined by varying the requirements on number of hits, distance of
closest approach and polar angle determination. As in the \piopio{}
case, the uncertainty on the modeling of Bose-Einstein correlations in
the Monte Carlo generator is obtained by using the control sample
\jetsetnobe{} in the analysis.  Finally, the binning in $Q$ is
varied.  The systematic uncertainties are attributed as in the
\piopio{} case, and are summarized in Table~\ref{tab:cpisyst}.  The
total systematic uncertainty is calculated as the quadratic sum of
these uncertainties.

\subsection*{Comparison}

The final values for the strength of the correlation $\lambda$ and the
corresponding radii of the boson sources $R$ are given in
Table~\ref{tab:final}.

Due to the lower efficiency of the \piopio{} selection, the
statistical uncertainty on the \piopio{} result is larger than the
statistical uncertainty on the \pipmpipm{} result. Within these
uncertainties, the data indicate both a weaker correlation and a
smaller source radius for \piopio{}. The weakness of the \piopio{}
correlation can be partly explained by the bigger contribution of
resonance decays to the $Q$ spectrum.  The difference of the source
sizes is
\begin{equation}
R_{\pm\pm} - R_{00} = 0.150 \pm 0.075 \,(\text{stat.})\,
                            \pm 0.068 \,(\text{syst.}) \fm,
\end{equation}
where $R_{\pm\pm}$ and $R_{00}$ indicate the value of $R$ for
\pipmpipm{} and \piopio{}, respectively. In this difference, the
systematic uncertainties due to the modeling of Bose-Einstein
correlations and the binning in $Q$ are taken to be correlated between
the two samples. The smaller radius found for \piopio{} is in
qualitative agreement with the predictions of the string model.

\section*{Acknowledgements}

We wish to express our gratitude to the CERN accelerator divisions for
the excellent performance of the LEP machine.  We acknowledge the
contributions of the engineers and technicians who have participated
in the construction and maintenance of this experiment.

\newpage


\newpage
\typeout{   }     
\typeout{Using author list for paper 242 -- ? }
\typeout{$Modified: Jul 31 2001 by smele $}
\typeout{!!!!  This should only be used with document option a4p!!!!}
\typeout{   }
%
%
%
%
%
%

\newcount\tutecount  \tutecount=0
\def\tutenum#1{\global\advance\tutecount by 1 \xdef#1{\the\tutecount}}
\def\tute#1{$^{#1}$}
\tutenum\aachen            
\tutenum\nikhef            
\tutenum\mich              
\tutenum\lapp              
\tutenum\basel             
\tutenum\lsu               
\tutenum\beijing           
\tutenum\berlin            
\tutenum\bologna           
\tutenum\tata              
\tutenum\ne                
\tutenum\bucharest         
\tutenum\budapest          
\tutenum\mit               
\tutenum\panjab            
\tutenum\debrecen          
\tutenum\florence          
\tutenum\cern              
\tutenum\wl                
\tutenum\geneva            
\tutenum\hefei             
\tutenum\lausanne          
\tutenum\lyon              
\tutenum\madrid            
\tutenum\florida           
\tutenum\milan             
\tutenum\moscow            
\tutenum\naples            
\tutenum\cyprus            
\tutenum\nymegen           
\tutenum\caltech           
\tutenum\perugia           
\tutenum\peters            
\tutenum\cmu               
\tutenum\potenza           
\tutenum\prince            
\tutenum\riverside         
\tutenum\rome              
\tutenum\salerno           
\tutenum\ucsd              
\tutenum\sofia             
\tutenum\korea             
\tutenum\utrecht           
\tutenum\purdue            
\tutenum\psinst            
\tutenum\zeuthen           
\tutenum\eth               
\tutenum\hamburg           
\tutenum\taiwan            
\tutenum\tsinghua          

{
\parskip=0pt
\noindent
{\bf The L3 Collaboration:}
\ifx\selectfont\undefined
 \baselineskip=10.8pt
 \baselineskip\baselinestretch\baselineskip
 \normalbaselineskip\baselineskip
 \ixpt
\else
 \fontsize{9}{10.8pt}\selectfont
\fi
\medskip
\tolerance=10000
\hbadness=5000
\raggedright
\hsize=162truemm\hoffset=0mm
\def\r{\rlap,}
\noindent

P.Achard\r\tute\geneva\ 
O.Adriani\r\tute{\florence}\ 
M.Aguilar-Benitez\r\tute\madrid\ 
J.Alcaraz\r\tute{\madrid,\cern}\ 
G.Alemanni\r\tute\lausanne\
J.Allaby\r\tute\cern\
A.Aloisio\r\tute\naples\ 
M.G.Alviggi\r\tute\naples\
H.Anderhub\r\tute\eth\ 
V.P.Andreev\r\tute{\lsu,\peters}\
% ...
\newpage
\begin{list}{A}{\itemsep=0pt plus 0pt minus 0pt\parsep=0pt plus 0pt minus 0pt
                \topsep=0pt plus 0pt minus 0pt}
\item[\aachen]
 I. Physikalisches Institut, RWTH, D-52056 Aachen, FRG$^{\S}$\\
 III. Physikalisches Institut, RWTH, D-52056 Aachen, FRG$^{\S}$
\item[\nikhef] National Institute for High Energy Physics, NIKHEF, 
     and University of Amsterdam, NL-1009 DB Amsterdam, The Netherlands
\item[\mich] University of Michigan, Ann Arbor, MI 48109, USA
\item[\lapp] Laboratoire d'Annecy-le-Vieux de Physique des Particules, 
     LAPP,IN2P3-CNRS, BP 110, F-74941 Annecy-le-Vieux CEDEX, France
\item[\basel] Institute of Physics, University of Basel, CH-4056 Basel,
     Switzerland
\item[\lsu] Louisiana State University, Baton Rouge, LA 70803, USA
\item[\beijing] Institute of High Energy Physics, IHEP, 
  100039 Beijing, China$^{\triangle}$ 
\item[\berlin] Humboldt University, D-10099 Berlin, FRG$^{\S}$
\item[\bologna] University of Bologna and INFN-Sezione di Bologna, 
     I-40126 Bologna, Italy
\item[\tata] Tata Institute of Fundamental Research, Mumbai (Bombay) 400 005, India
\item[\ne] Northeastern University, Boston, MA 02115, USA
\item[\bucharest] Institute of Atomic Physics and University of Bucharest,
     R-76900 Bucharest, Romania
\item[\budapest] Central Research Institute for Physics of the 
     Hungarian Academy of Sciences, H-1525 Budapest 114, Hungary$^{\ddag}$
\item[\mit] Massachusetts Institute of Technology, Cambridge, MA 02139, USA
\item[\panjab] Panjab University, Chandigarh 160 014, India.
\item[\debrecen] KLTE-ATOMKI, H-4010 Debrecen, Hungary$^\P$
\item[\florence] INFN Sezione di Firenze and University of Florence, 
     I-50125 Florence, Italy
\item[\cern] European Laboratory for Particle Physics, CERN, 
     CH-1211 Geneva 23, Switzerland
\item[\wl] World Laboratory, FBLJA  Project, CH-1211 Geneva 23, Switzerland
\item[\geneva] University of Geneva, CH-1211 Geneva 4, Switzerland
\item[\hefei] Chinese University of Science and Technology, USTC,
      Hefei, Anhui 230 029, China$^{\triangle}$
\item[\lausanne] University of Lausanne, CH-1015 Lausanne, Switzerland
\item[\lyon] Institut de Physique Nucl\'eaire de Lyon, 
     IN2P3-CNRS,Universit\'e Claude Bernard, 
     F-69622 Villeurbanne, France
\item[\madrid] Centro de Investigaciones Energ{\'e}ticas, 
     Medioambientales y Tecnolog{\'\i}cas, CIEMAT, E-28040 Madrid,
     Spain${\flat}$ 
\item[\florida] Florida Institute of Technology, Melbourne, FL 32901, USA
\item[\milan] INFN-Sezione di Milano, I-20133 Milan, Italy
\item[\moscow] Institute of Theoretical and Experimental Physics, ITEP, 
     Moscow, Russia
\item[\naples] INFN-Sezione di Napoli and University of Naples, 
     I-80125 Naples, Italy
\item[\cyprus] Department of Physics, University of Cyprus,
     Nicosia, Cyprus
\item[\nymegen] University of Nijmegen and NIKHEF, 
     NL-6525 ED Nijmegen, The Netherlands
\item[\caltech] California Institute of Technology, Pasadena, CA 91125, USA
\item[\perugia] INFN-Sezione di Perugia and Universit\`a Degli 
     Studi di Perugia, I-06100 Perugia, Italy   
\item[\peters] Nuclear Physics Institute, St. Petersburg, Russia
\item[\cmu] Carnegie Mellon University, Pittsburgh, PA 15213, USA
\item[\potenza] INFN-Sezione di Napoli and University of Potenza, 
     I-85100 Potenza, Italy
\item[\prince] Princeton University, Princeton, NJ 08544, USA
\item[\riverside] University of Californa, Riverside, CA 92521, USA
\item[\rome] INFN-Sezione di Roma and University of Rome, ``La Sapienza",
     I-00185 Rome, Italy
\item[\salerno] University and INFN, Salerno, I-84100 Salerno, Italy
\item[\ucsd] University of California, San Diego, CA 92093, USA
\item[\sofia] Bulgarian Academy of Sciences, Central Lab.~of 
     Mechatronics and Instrumentation, BU-1113 Sofia, Bulgaria
\item[\korea]  The Center for High Energy Physics, 
     Kyungpook National University, 702-701 Taegu, Republic of Korea
\item[\utrecht] Utrecht University and NIKHEF, NL-3584 CB Utrecht, 
     The Netherlands
\item[\purdue] Purdue University, West Lafayette, IN 47907, USA
\item[\psinst] Paul Scherrer Institut, PSI, CH-5232 Villigen, Switzerland
\item[\zeuthen] DESY, D-15738 Zeuthen, 
     FRG
\item[\eth] Eidgen\"ossische Technische Hochschule, ETH Z\"urich,
     CH-8093 Z\"urich, Switzerland
\item[\hamburg] University of Hamburg, D-22761 Hamburg, FRG
\item[\taiwan] National Central University, Chung-Li, Taiwan, China
\item[\tsinghua] Department of Physics, National Tsing Hua University,
      Taiwan, China
\item[\S]  Supported by the German Bundesministerium 
        f\"ur Bildung, Wissenschaft, Forschung und Technologie
\item[\ddag] Supported by the Hungarian OTKA fund under contract
numbers T019181, F023259 and T024011.
\item[\P] Also supported by the Hungarian OTKA fund under contract
  number T026178.
\item[$\flat$] Supported also by the Comisi\'on Interministerial de Ciencia y 
        Tecnolog{\'\i}a.
\item[$\sharp$] Also supported by CONICET and Universidad Nacional de La Plata,
        CC 67, 1900 La Plata, Argentina.
\item[$\triangle$] Supported by the National Natural Science
  Foundation of China.
\end{list}
}
\vfill


\newpage

\begin{table}[hbt]
\begin{center}
\begin{tabular}{|l|ccc|}
\hline
Source
& $\Delta\lambda$
& $\Delta R$ (\!\!\fm)
& $\Delta\alpha$ \\
\hline
Photon selection &
0.020 & 0.052 & 0.009 
\\
Mass window &
0.011 & 0.008 & 0.007
\\
2D fit range &
0.056 & 0.019 & 0.036
\\
MC modeling &
0.037 & 0.035 & 0.020
\\
Data-MC agreement &
0.012 & 0.018 & 0.034
\\
$Q$-binning &
0.004 & 0.013 & 0.003
\\
\hline
Total &
0.072 & 0.070 & 0.055 
\\
\hline
\end{tabular}
\end{center}
 \caption{ Systematic uncertainties on $\lambda$, $R$ and
   $\alpha$ for the \piopio{} data sample.
   \label{tab:pi0syst} }
\end{table}


\begin{table}[hbt]
\begin{center}
\begin{tabular}{|l|ccc|}
\hline
Source
& $\Delta\lambda$
& $\Delta R$ (\!\!\fm)
& $\Delta\alpha$ \\
\hline
Track selection &
0.011 & 0.009 & 0.009
\\
MC modeling &
0.022 & 0.003 & 0.002
\\
$Q$-binning &
0.001 & 0.001 & 0.001 
\\
\hline
Total &
0.025 & 0.010 & 0.009
\\
\hline
\end{tabular}
\end{center}
 \caption{ Systematic uncertainties on $\lambda$, $R$ and
   $\alpha$ for the \pipmpipm{} data sample.
   \label{tab:cpisyst} }
\end{table}


\renewcommand{\arraystretch}{1.2}
\begin{table}[hbt]
\begin{center}
\begin{tabular}{|l|ccc|}
\hline
Sample & $\lambda$ & $R$ (\!\!\fm) & $\alpha$ \\
\hline
\piopio{} &
$0.155 \pm 0.054 \pm 0.072 $ &
$0.309 \pm 0.074 \pm 0.070 $ & 
$0.021 \pm 0.034 \pm 0.055 $
\\
\pipmpipm{} &
$0.286 \pm 0.008 \pm 0.025 $ &
$0.459 \pm 0.010 \pm 0.010 $ &
$0.015 \pm 0.003 \pm 0.009 $
\\
\hline
\end{tabular}
\end{center}
\caption{ Values for $\lambda$, $R$ and $\alpha$, for both
  the \piopio{} and the \pipmpipm{} data samples. The first uncertainty is
  statistical, the second systematic. 
  \label{tab:final} }
\end{table}
\renewcommand{\arraystretch}{1.}

\newpage

\newpage

\newpage

\newpage

\end{document}